# Electron-Hole Pair Generation in SiC High-Temperature Alpha Particle Detectors

Timothy R Garcia[1], Ashutosh Kumar[2], Benjamin Reinke[1], Thomas E. Blue[1], and Wolfgang Windl[2]

[1]Nuclear Engineering Program, The Ohio State University, Columbus, OH, 43210, USA
[2]Materials Science and Engineering Department, The Ohio State University, Columbus, OH, 43210, USA

*Abstract*— We demonstrate alpha-particle detection in an n-type 4H-SiC Schottky diode detector up to an unprecedented temperature of 500 °C using an Am-241 disc source. The measured spectra were used to calculate the electron-hole pair creation energy in 4H-SiC and its non-bandgap contribution, which are both found to decrease with increasing temperature. The full width at half maximum (FWHM) of the measured alpha-energy peaks was found to increase exponentially with temperature due to an exponential increase of leakage current. For our measurement system, above 300 °C, where the leakage current was $10^{-6}$ A, this increase exceeded the FWHM at room temperature.

## I. INTRODUCTION

Alpha-particle spectroscopy is a technique that is commonly used to detect and quantify the presence of alpha-emitters. Since alpha energies are often very similar in value, high-resolution devices are desirable because they allow discrimination between the different alpha-emitting radioisotopes. State-of-the art solid-state alpha particle detectors are commonly made from silicon[1]. However, Si detectors are limited to operating at temperatures barely above room temperature, due to the low band gap of silicon[2]. This has made alpha-particle detection at "extreme" temperatures, to date, impossible with silicon-based devices.

Functioning alpha-particle detectors are highly desirable for high-temperature applications, such as *in-situ* measurements of actinide concentrations for safeguarding the pyrochemical reprocessing of spent nuclear fuel, which is performed at a minimum temperature of 500 $^{o}C$[3]. In order to perform measurements at high temperature with a solid-state detector, a wide-band gap material must be used. 4H-SiC is one of the most obvious candidate materials for high temperature solid-state detector applications because of its large band gap, reported radiation hardness[4], high charge carrier mobility[5], and successful demonstration as a room-temperature alpha particle detector[4,6]. However, alpha-particle detection in silicon carbide has only been reported to date for temperatures below 89 $^{o}C$ for Schottky[6] and 375 $^{o}C$ for pn-type diode detectors[4]. In both cases the depletion width was thinner than the range of the alpha particles, resulting in incomplete charge collection. The ability of alpha particle detectors to distinguish between alpha particles of different energies relies on complete charge collection; so these experiments, with incomplete charge collection, did not yield results which are conclusive regarding energy discrimination.

In this letter we report, for the first time from room temperature to 500 °C, values of $\varepsilon_{4H\text{-}SiC}$, the average energy required to yield an electron-hole pair (ehp) in 4H-SiC. The values were calculated from Am-241 alpha particle energy spectra measured by a 4H-SiC Schottky diode detector, which we fabricated and biased to ensure that the measurements were performed with full charge collection. Finally, the relationship between the temperature-dependent leakage current and full width at half maximum (FWHM) of the charge collection Gaussian peaks is shown to be linear above 300 °C.

## II. BACKGROUND / EXPERIMENTAL METHODS

The detectors used were built on an n-type 4H-SiC wafer from Cree, Inc., consisting of a ~300 μm thick bulk 4H-SiC layer with ~$1\times10^{18}$ $cm^{-3}$ nitrogen doping, and a 21 μm thick epitaxial layer with ~$5.3\times10^{14}$ $cm^{-3}$ nitrogen doping. This wafer was chosen because the epitaxial layer was thicker than the range of Am-241 alpha particles in SiC, which is about 18 μm. The wafer was diced into square samples, with side lengths of 8 mm, by Kadco ceramics. The entire backside (bulk layer) was metallized with nickel and annealed at 950 °C for 30 seconds to form an Ohmic contact. The epitaxial layer was shadow-masked with circular contacts of 3 mm diameter, metallized with nickel, titanium, and gold, and annealed at 650 °C for 30 seconds to form a Schottky contact.

By applying a reverse-bias voltage of about 200 V across the contacts a depletion region, the active volume for radiation detection, was maintained throughout the entire epitaxial layer allowing for full charge collection. When a charged particle interacts with this active volume, a cloud of bound charge carriers, in this case electron-hole pairs, are excited, i.e. freed to move, and can be collected at the electrical contacts. The total number of electron-hole pairs collected ($N$) can be related to the energy of the incident alpha particle through[2]

$$N = \frac{E_\alpha}{\varepsilon_{\text{4H-SiC}}}, \tag{1}$$

where $E_\alpha$ is the energy of the alpha particle, and $\varepsilon_{4H\text{-}SiC}$ is as it was defined above. In order to measure the energy of alpha particles accurately at varying temperatures, knowledge of the temperature dependence of $\varepsilon_{4H\text{-}SiC}$ is necessary. The room-temperature value of $\varepsilon_{4H\text{-}SiC}$ for direct electron irradiation has been measured to equal 5.05 eV/ehp[7]; while measurements of $\varepsilon_{4H\text{-}SiC}$ for light and heavy ions result in larger values of 7.28 eV/ehp[8], or 7.8 eV/ehp[9].

In order to test the detectors at high temperature, a vacuum bell jar test chamber was fabricated containing an alpha particle disc source and an Inconel substrate heater from Blue Wave Semiconductors (Fig. 1). The detector studied was held to the heater using a steel clip to maintain good thermal and electrical contact. Two micropositioners with tungsten probe tips were used as high-temperature electrical connections. Depending upon the temperature of the detector, the pulses were amplified by the appropriate one of two Ortec 142B preamplifiers. The output pulses from the preamplifier were processed to form pulse height spectra using a Canberra DSA 2000 digital signal analyzer. The leakage current at 200 V was measured using a Keithley 2410 sourcemeter. Leakage current increased strongly with temperature, which caused an increase of the voltage drop across the bias resistor in the preamp used, and a corresponding drop in voltage across the detector itself. At each temperature, the applied voltage was increased to fully compensate for this drop in voltage. An unmodified Ortec 142B preamp with 100 MΩ bias resistor was used from room temperature to 200 °C, at which point the applied voltage was greater than the voltage limit of the preamp. A modified Ortec 142B preamp with 1 MΩ bias resistor was used from room temperature to 500 °C.

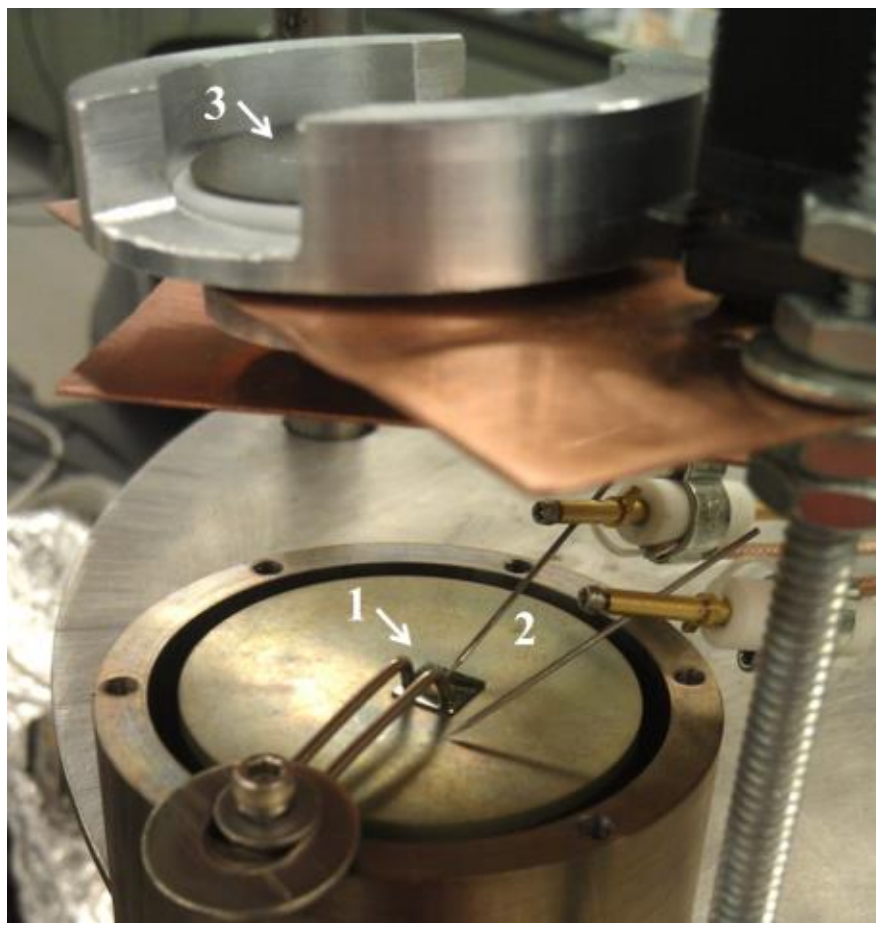

Figure 1: Experimental setup for high-temperature alpha irradiation of a 4H-SiC Schottky diode detector. The detector (1) is clipped onto the heater surface. (2) indicates the two tungsten probe micropositioners used as electrical contacts, and (3) the back surface of the Am-241 electroplated alpha-particle disc source.

To measure $\varepsilon_{4H\text{-}SiC}$, we adopted the method suggested in Ref. 8. A conversion factor was measured that converted from channel in the DSA 2000 to number of charges input to the preamp. This conversion factor was measured using a known charge injected into the preamplifier by a pulse generator, and monitoring the corresponding channel in the DSA 2000. Then, the 4H-SiC detector was irradiated by Am-241 alpha particles. The value of $\varepsilon_{4H\text{-}SiC}$ was then calculated assuming nominal values for the alpha particles energies, and converting from channel to charge using the aforementioned conversion factor.

Specifically, an Ortec 419 pulse generator was used with its supplied charge terminator, which had a measured capacitance (C) of 2.37 pF. The pulse generator output was adjusted so that the spectrum recorded by the DSA 2000 had the centroid of the pulser's peak ($n_p$) in approximately the same channel ($n_{E_1}$) as the centroid of the $E_1$ = 5.388 MeV alpha particle peak for the Am-241 spectrum at room-temperature. The voltage pulse height (V) was measured using an oscilloscope. Then, the total number of electron-hole pairs collected, $N$, was calculated using

equation (2)

$$N = \frac{CV}{q} \tag{2}$$

where $q$ is the elementary charge and $C$ is the measured capacitance. Next, a conversion factor ($c_c$) from charge carriers in the input of the preamplifier to channel number in the DSA 2000 was calculated using $N$ from equation (2) in equation (3), as

$$c_c = \frac{n_p}{N}. \tag{3}$$

An Eberline Services electroplated 1.88 μCi Am-241 disc source was used for the alpha energy spectra measurements. Vacuum was maintained below 30 mTorr to minimize the effect of alpha-particle air interactions on the measured energies. The energy loss due to finite Am-241 thickness in the disc source was measured using an Ortec Si ULTRA alpha particle detector at room temperature by comparing $E_\alpha$ as calculated using equations (1) and (2) and a common value for $\varepsilon_{Si}$ of 3.6 eV/ehp, with $E_\alpha$ = 5.388 MeV[2]. This energy loss was determined to be about 20 keV. For temperature-dependent measurements using the SiC-detector, the detector temperature was varied between room temperature and 500 °C in increments of 50 °C. To eliminate electrical noise from the heater, alpha spectra were only measured when the heater was off, with a heating cycle time that maintained a swing of no more than ±5 °C about each measurement temperature.

The measured spectra were processed as described below. First, using the DSA 2000, the alpha particle energies were binned into ~2.9 keV wide channels for the SiC-detector spectra. Then, the measured spectra, as a function of channel number ($n$), $S(n)$, were fit as a sum of three Gaussian curves for the three prominent alpha particle peak energies ($E_1$ = 5.388 MeV, $E_2$ = 5.443 MeV, and $E_3$ = 5.486 MeV). These energies contribute more than 99.5% of the alpha particles emitted by Am-241. The functional form to which the spectra were fit is

$$S(n) = k \sum_i A_i \exp\left(-4 \ln 2 \left\{n - \frac{E_i}{\Delta E}\right\}^2 / \mathrm{FWHM}^2\right), \tag{4}$$

where $A_1 = 1.7$, $A_2 = 13.1$, and $A_3 = 84.8$ are the relative intensities for Am-241 $\alpha$ emission[10]; and the fitting parameters are $\Delta E$ (channel width); FWHM (full width at half max in number of channels); and $k$ (magnitude of the measured spectra). The parameters $\Delta E$ and FWHM are functions of temperature. Lastly, $\varepsilon_{\text{4H-SiC}}$ was calculated using

$$\varepsilon_{\text{4H-SiC}} = c_c \Delta E. \tag{5}$$

III. RESULTS AND DISCUSSION

Using $\Delta E$ from eqn. (4) in eqn. (5) yielded a room-temperature value for $\varepsilon_{\text{4H-SiC}}$ of 7.82 ± 0.02 eV/ehp. This value matches the result of Nava[9] within one standard deviation, but is larger than the value found by Chaudhuri[8] by about 0.5 eV.

Figure 2 shows two examples of measured spectra from the same detector measured at 23 °C and 450 °C. One can see that both the centroids of the Gaussian peaks and the FWHM of the Gaussian peaks are larger at higher temperatures.

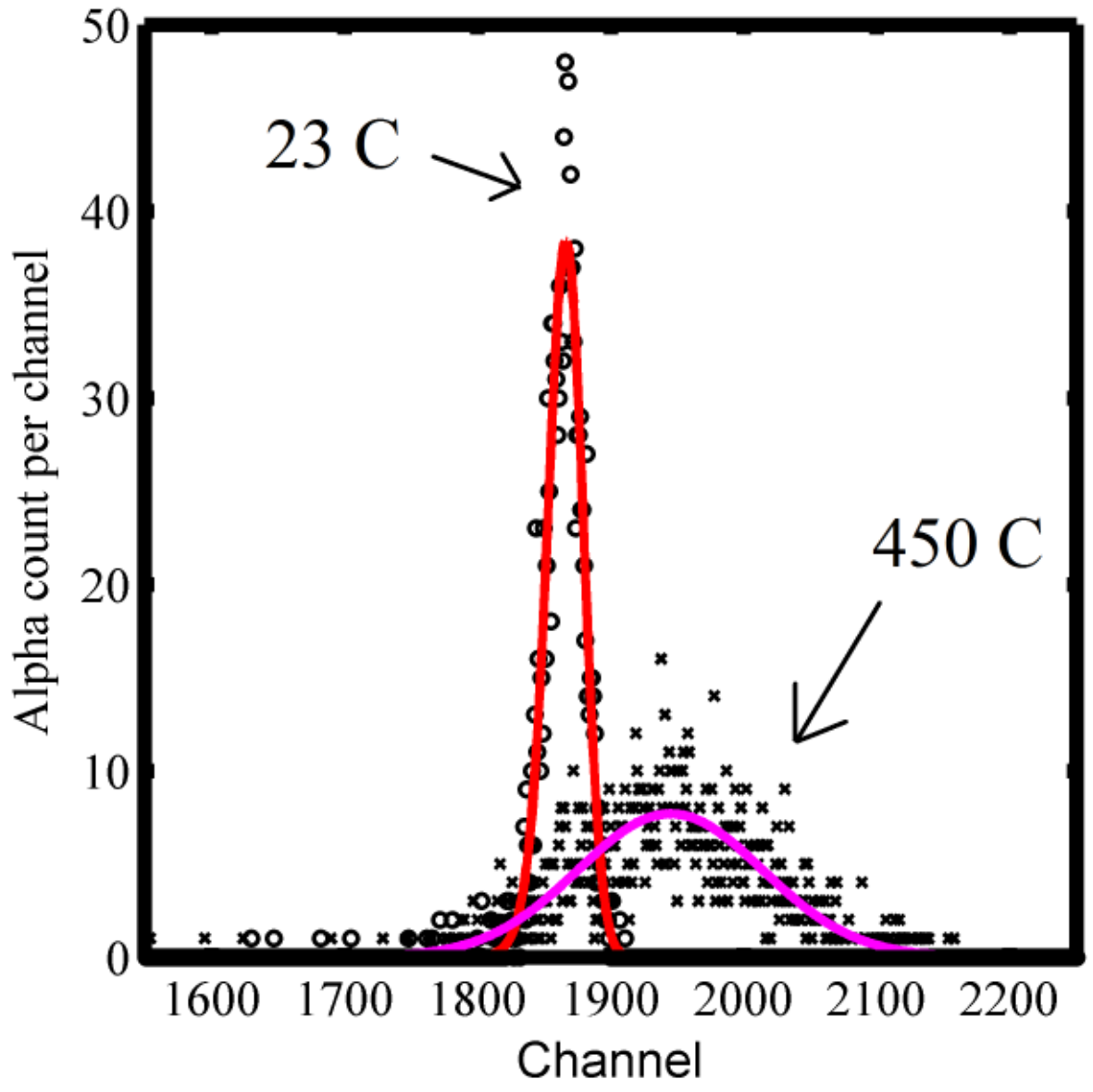

Figure 2: Am-241 disc source alpha-particle spectra measured with a 3 mm-diameter 4H-SiC Schottky diode detector at 23 °C and 450 °C. Due to a relatively large FWHM at room temperature, the three major peaks of Am-241 are indistinguishable at room temperature. As temperature increases, there is an increase in both centroid and FWHM of the alpha peaks.

Consistent with the movement of the centroids, $\varepsilon_{4H\text{-}SiC}$ decreases with increasing temperature (Figure 3).

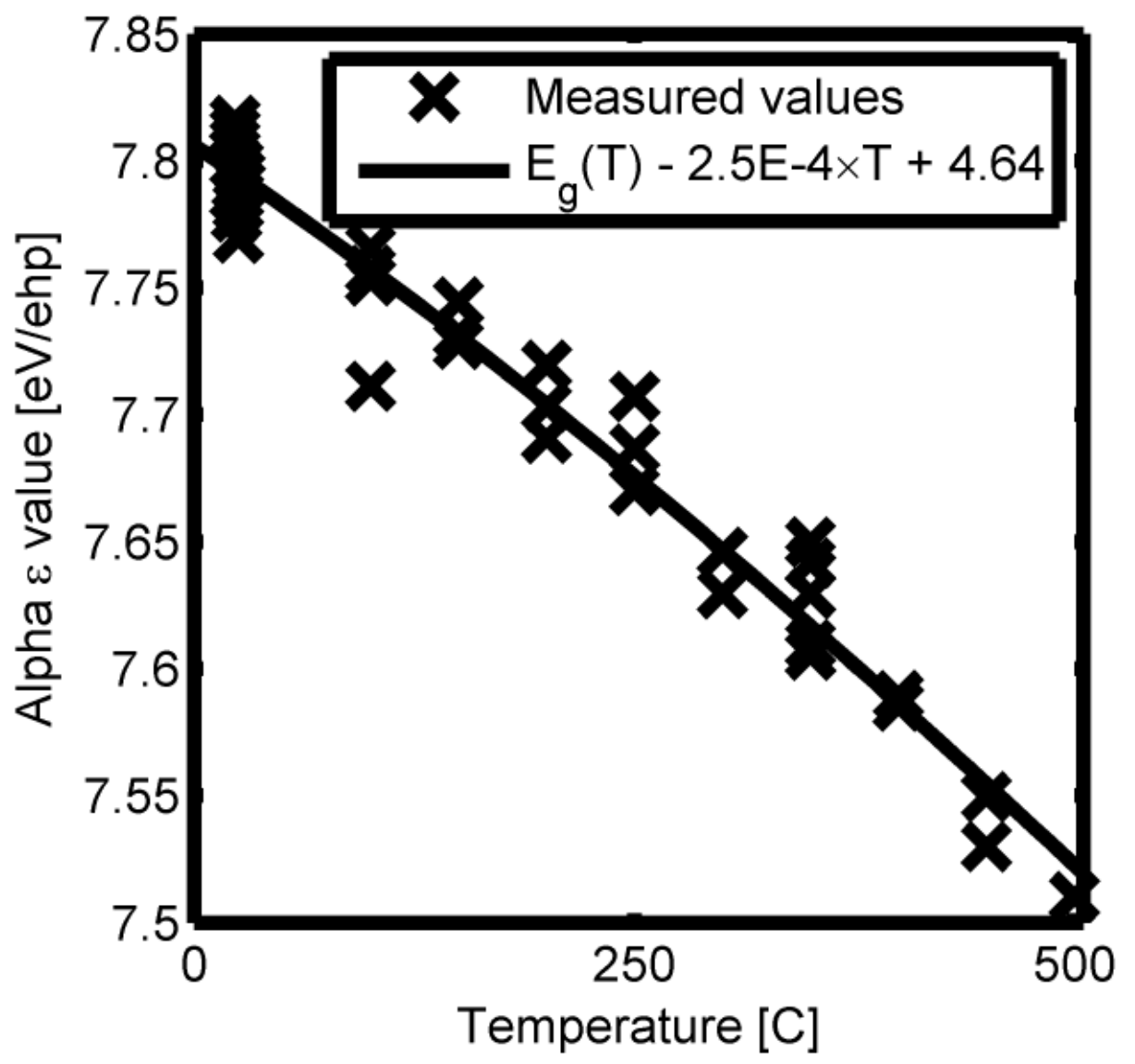

**Figure 3:** Measured electron-hole generation energy in 4H-SiC as a function of temperature in $^{o}$C, determined from Am-241 peak positions. X's indicate experimental values. The line is a fit as a function of temperature with the equation shown in the legend with temperature in $^{o}$C and with the experimental temperature dependence of the band gap $\boldsymbol{E_g}$ as given in equation (7).

To understand this temperature dependence, we start with Klein's model[11], which states that

$$\varepsilon_{\text{4H-SiC}} = E_g + \langle E_K \rangle + \langle E_R \rangle, \tag{6}$$

where $\langle E_K \rangle$ is the average kinetic energy remaining on electrons and holes when their energy is too low for them to create additional electron-hole pairs, $\langle E_R \rangle$ is the average energy loss due to energy transfer to optical phonons, and $E_g$ is the band gap energy. The decrease of the band gap energy with temperature in 4H-SiC has been described by[5]

$$E_g(\theta) = 3.265 - 6.5 \times 10^{-4} \frac{(\theta+273)^2}{\theta+1573} \text{ [eV]}, \tag{7}$$

where $\theta$ is in units of $^{o}$C. According to equation (7), $E_g$ decreases by 0.15 eV when the temperature is increased from 23 $^{o}$C to 500 $^{o}$C. However, the measured decrease in $\varepsilon_{4H\text{-}SiC}$ over the same temperature range is nearly twice as large, with a value of about 0.27 eV. This means that the sum of $\langle E_K \rangle + \langle E_R \rangle$ must decrease by ~0.12 eV to account for the decrease in $\varepsilon_{4H\text{-}SiC}$.

To find the temperature dependence of the sum of $\langle E_K \rangle + \langle E_R \rangle$, $\varepsilon_{4H\text{-}SiC}(\theta)$ was fit with a summation of the band gap function from equation (7) and a linear function,

$$\varepsilon_{4H\text{-}SiC} = E_g(\theta) - a\theta + b. \quad (8)$$

Figure 3 shows the resulting fit. The coefficients $a$ = 2.5x10$^{-4}$ ± 0.3x10$^{-4}$ eV/°C and $b$ = 4.64±0.02 eV are reported in the legend. With $a$ = 2.5E-4 eV/°C and $b$ = 4.64 eV, the sum $\langle E_K \rangle + \langle E_R \rangle$ decreases linearly from 4.57 eV at room temperature to 4.45 eV at 500 $^{o}$C.

We can rationalize that $\langle E_K \rangle + \langle E_R \rangle$ decreases with increasing temperature given the physics that describe $\langle E_K \rangle$ and $\langle E_R \rangle$. Previous work has suggested $\langle E_K \rangle$ is directly proportional to $E_g$, e.g. $\langle E_K \rangle = \frac{9}{5} E_g$[11]. Therefore, if $E_g$ decreases with an increase of temperature, then $\langle E_K \rangle$ decreases as well. Klein stated that $\langle E_R \rangle$ can be estimated by $\langle r \rangle \langle \hbar\omega_R \rangle$, where $\langle r \rangle$ is the average number of phonons created with an average energy $\langle \hbar\omega_R \rangle$[11]. For 4H-SiC, $\langle \hbar\omega_R \rangle = 104$ meV. According to Bose-Einstein statistics, the probability of finding an available 104 meV optical phonon state in 4H-SiC decreases from about 98% at room temperature to about 72% at 500 °C. This effect reduces $\langle r \rangle$ in direct proportion. Since the temperature dependence of phonons in SiC has been found to be negligible[12], the product $\langle r \rangle \langle \hbar\omega_R \rangle = \langle E_R \rangle$ thus decreases with an increase of temperature.

Charge carrier recombination was not included in the model for $\varepsilon_{4H\text{-}SiC}$ given by equation (6). If recombination played a dominant role in charge carrier collection, then the terms $-a\theta + b$ in equation (8) would increase. Since these terms have been observed to decrease, we can state that charger carrier recombination does not play a dominant role in these detectors.

With the system properly calibrated to Am-241 alpha particle energy at various temperatures, the temperature dependence of FWHM can be investigated in terms of energy rather than channel. Figure 4 shows a semi-logarithmic plot of the FWHM of the detector, in units of keV, versus the temperature of the detector. Below 300 °C, there is no significant degradation in alpha energy resolution. Above 300 °C, the FWHM increases approximately linearly on a semi-logarithmic scale, i.e. exponentially with increasing temperature. Figure 4 also shows that, for all temperatures, the leakage current for this detector follows an exponential dependence. This strongly suggests that the exponential increase in FWHM above 300 $^{o}$C observed in Figure 4 is caused by the exponential increase in leakage current, also shown in Figure 4. The temperature 300 $^{o}$C corresponds to a leakage current of ~$10^{-6}$ A, indicating a leakage current threshold under which no degradation in FWHM is observed. This also suggests that for leakage currents smaller than ~$10^{-6}$ A, the FWHM for this detector is dominated by other mechanisms of resolution degradation, such as thermal (Johnson) noise originating in the preamplifier, as has been previously suggested for diode detector systems in general[2], and/or alpha particle energy straggling in the Schottky metal contact.. A relatively large amount of alpha particle energy straggling is possible in this layer, since, to ensure mechanical robustness, the Schottky metal contact was grown to be relatively thick at 120 nm,. Because of this, and other temperature-invariant contributions to FWHM, the room temperature FWHM is greater than the measurements by others[8,13].

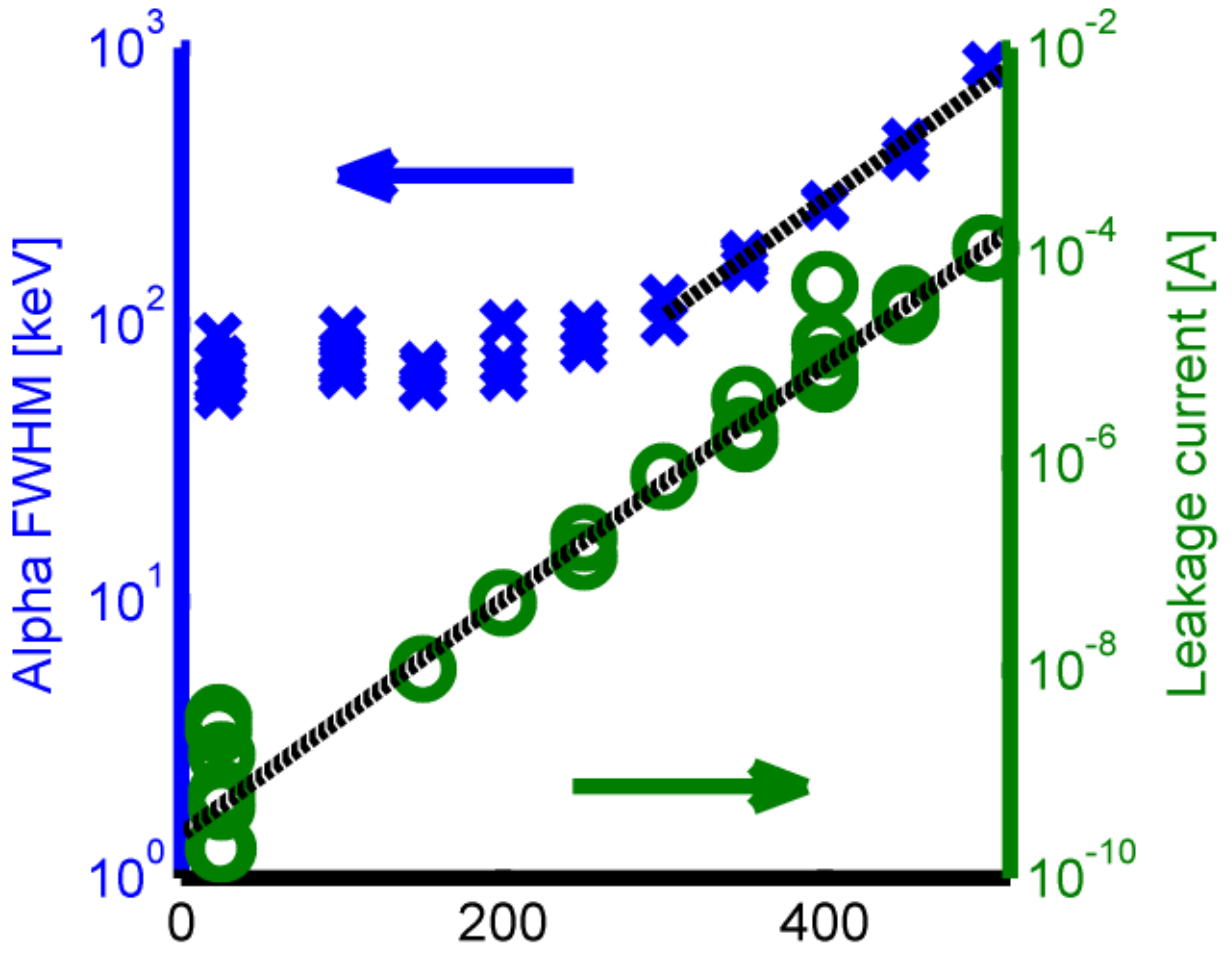

**Figure 4: Semilog plots of Am-241 alpha particle energy spectra FWHM in units of keV (left) and leakage current at 200 V reverse bias in units of amps (right) versus temperature for a 3 mm-diameter 4H-SiC detector.**

## IV. CONCLUSIONS

In summary, we have demonstrated the ability to reliably detect Am-241 alpha particle energy spectra up to a previously unachieved temperature of 500 °C with 4H-SiC Schottky diode detectors. We found an increase of centroid as well as FWHM with temperature. Analysis of our data allowed us to quantify the temperature-dependence of the electron-hole pair creation energy. Analysis of the electron-hole pair creation energy showed that the centroid increase was due to a decrease in both 4H-SiC band gap and non-ionizing energy loss, and that carrier recombination seems to play, at best, a minor role. We also found that the FWHM increase seems to be dominated by leakage current. Since we did not observe degradation in detector parameters based on repeated heating and cooling of this detector (13 cycles), such detectors may be appropriate for high temperature operation for extended periods, although further studies are necessary to confirm this. With an optimized Schottky contact, a 4H-SiC detector may be able to operate even above 500 °C and with a lower leakage current (i.e. smaller FWHM) at 500 °C. Our results suggest that in order to pursue research into high-temperature alpha particle detectors, an emphasis should be put on minimizing the leakage current at high temperature. Now, experiments which test these detectors at a combination of high temperature and high doses of damaging radiation are needed in order to identify their effects on charge carrier collection separately and in combination, and to determine the ultimate limits of operating parameters for these detectors.

**Acknowledgments.** This research was performed using funding received from the DOE Office of Nuclear Energy's Nuclear Energy University Programs under Grant No. 09-842. This work was supported in part by an allocation of computing time from the Ohio Supercomputer Center. T.G. would also like to acknowledge the NASA Graduate Student Research Program for their support.

## V. References

[1] E. García-Toraño, Appl. Radiat. Isot. **64**, 1273 (2006).
[2] G.F. Knoll, *G. F. Knoll's Radiation Detection 3rd (Third) edition(Radiation Detection and Measurement [Hardcover])(2000)*, 3 edition edition (Wiley, n.d.).
[3] C.W. Johnson, M.L. Dunzik-Gougar, and S.X. Li, Trans.-Am. Nucl. Soc. **95**, 129 (2006).
[4] E.V. Kalinina, A.M. Ivanov, N.B. Strokan, and A.A. Lebedev, Semicond. Sci. Technol. **26**, 045001 (2011).
[5] M.E. Levinshtein, S.L. Rumyantsev, and M.S. Shur, *Properties of Advanced Semiconductor Materials: GaN, AIN, InN, BN, SiC, SiGe* (John Wiley & Sons, 2001).
[6] F.H. Ruddy, A.R. Dulloo, J.G. Seidel, S. Seshadri, and L.B. Rowland, IEEE Trans. Nucl. Sci. **45**, 536 (1998).

[7] M.V.S. Chandrashekhar, C.I. Thomas, and M.G. Spencer, Appl. Phys. Lett. **89**, 042113 (2006).
[8] S.K. Chaudhuri, K.J. Zavalla, and K.C. Mandal, Appl. Phys. Lett. **102**, 031109 (2013).
[9] F. Nava, G. Wagner, C. Lanzieri, P. Vanni, and E. Vittone, Nucl. Instruments Methods Phys. Res. Sect. Accel. Spectrometers Detect. Assoc. Equip. **510**, 273 (2003).
[10] M.S. Basunia, Nucl. Data Sheets **107**, 2323 (2006).
[11] C.A. Klein, J. Appl. Phys. **39**, 2029 (1968).
[12] F.H. Ruddy, J.G. Seidel, Haoqian Chen, A.R. Dulloo, and Sei-Hyung Ryu, IEEE Trans. Nucl. Sci. **53**, 1713 (2006).